\documentclass[showpacs,twocolumn,nofootinbib,nobibnotes,pra]{revtex4}
\usepackage{amsfonts}
\usepackage{amsmath}
\usepackage{amssymb}
\usepackage{graphicx}

\input{tcilatex}

\begin{document}

\title{Quantum electrodynamics in a whispering gallery microcavity coated
with a polymer nanolayer}
\author{Yun-Feng Xiao$^{1}$}
\email{yfxiao@pku.edu.cn}
\author{Chang-Ling Zou$^{2}$}
\author{Peng Xue$^{3}$}
\author{Lixin Xiao$^{1}$}
\author{Yan Li$^{1}$}
\author{Chun-Hua Dong$^{2}$}
\author{Zheng-Fu Han$^{2}$}
\author{Qihuang Gong$^{1}$}
\email{qhgong@pku.edu.cn}
\affiliation{$^{1}$State Key Laboratory for Artificial Microstructure and Mesoscopic
Physics, School of Physics, Peking University, Beijing 100871, P. R. China.}
\affiliation{$^{2}$Key Lab of Quantum Information, CAS, University of Science and
Technology of China, Hefei 230026, P. R. China.}
\affiliation{$^{3}$Department of Physics, Southeast University, Nanjing 211189, P. R.
China}

\begin{abstract}
Quasi-TE and TM fundamental whispering gallery modes in a polymer coated
silica microtoroid are theoretically investigated, and demonstrated to
possess very high quality factors. The existence of nanometer thickness
layer not only reduces the cavity mode volume evidently, but also draws the
the maximal electric field's position of the mode to the outside of the
silica toroid where single quantum dots or nanocrystals are located. Both
effects result in a strongly enhanced coherent interaction between single
dipole (for example, single defect center in diamond crystal) and the
quantized cavity mode. Since the coated microtoroid is highly feasible and
robust in experiment, it may offer an excellent platform to study
strong-coupling cavity QED, quantum information and quantum computation.
\end{abstract}

\maketitle

\section{Introduction}

When single neutral atoms or quantum dots are strongly coupled to quantized
electromagnetic fields through dipole-interaction inside a high-$Q$ cavity,
it has long been a central paradigm for the study of open quantum systems
and plays a leading role in defining research goals. This coherent
interaction is referred to cavity quantum electrodynamics (QED), which
offers an almost ideal system for the generation of entangled states and
implementation of small-scale quantum information processors recently \cite%
{QED1,QED2}. The reason is that atoms are particularly well suited for
storing qubits in long-lived internal states, and photons are the best qubit
carrier for fast and reliable communication over long distances.

For cavity QED study, three representative optical microcavities have been
proposed and investigated up to the present \cite{vahala}: the conventional
Fabre-Perot(FP)-type cavities consisting of two concave dielectric mirrors
at a distance of tens of microns, the nanoscale optical cavities in photonic
crystal where a dot or line defect is introduced, and the dielectric optical
microcavities supporting whispering gallery modes (WGMs). Over past few
years, the strong coupling regime (i.e., the light-matter coherent
interaction strength exceeds both the cavity and dipole decay rates) has
been realized using either FP-type microcavities \cite{SC1,SC2,SC3,SC4} or
photonic crystal nanocavities \cite{SC5,SC6,SC7}. Experiments with single
atoms have been at the forefront of these advances with the use of optical
FP resonators. Nevertheless, the extremely technical challenges will be
typically involved in further improving the resonators and scaling to large
numbers of microcavities. Thus there is an increasing interest in the
development of alternative microcavity systems. WGMs in a
rotational-symmetry geometry exhibit exciting characteristic such as
extremely high quality factor, small mode volume, and excellent scalability 
\cite{buck,spillane}. As a result, this system has been of particular
interest to reach the strong light-matter coherent interaction. In
experiment, single cold caesiums coupled to a silica microtoroid \cite{aoki}%
, nanocrystalline quantum dots coupled to microsphere cavities \cite%
{thomas,park}, and single quantum dots coupled microdisks \cite%
{peter,painter} have been reported.

Except single etched quantum dots embedding in the microdisks, single cold
atoms or nanocrystalline quantum dots are locating on the outer surface of
cavity. In this case, only weak evanescent field of the cavity mode is used
to interact with atoms/quantum dots, and thus the coherent coupling strength
can not reach its maximum. In this paper, we study cavity QED in a silica
toroid microcavity coated with a high-refractive-index nanolayer. The idea
is at least twofold. First, with an increasing in nanolayer thickness, the
optical field of a WGM in the coated microtoroid moves to outside the silica
surface where quantum dots are located, which results in strongly enhancing
the coherent interaction strength between the WGM and coupled quantum dots.
Second, the high-refractive-index layer compresses the radial optical field
of the WGM, which may reduce the mode volume.

\begin{figure}[ptbh]
\centerline{\includegraphics[keepaspectratio=true,width=0.45%
\textwidth]{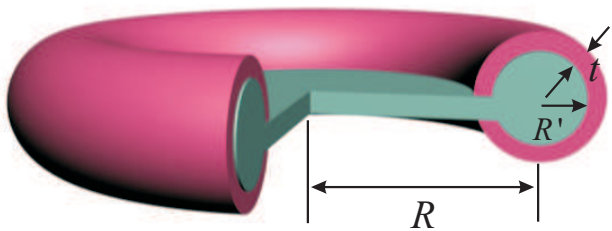}}
\caption{(Color online) Schematic illustration of a polymer coated silica
toroidal microcavity. The major and minor radii of silica toroid are $R$ and 
$R^{\prime}$, respectively. The coating thickness is $t$. The refractive
indexes of silica, polymer and air are $n_{1}=1.4564$, $n_{2}=1.59$, and $%
n_{3}=1$, respectively.}
\end{figure}

\section{Modes in a nanolayer-coated microtoroid}

The geometry of the proposed nanolayer-coated microcavity is shown in Fig.
1. A silica microtoroid with a refraction index $n_{1}$\ can be prepared on
a silicon chip through the standard photolithography technique and CO2 laser
reflow \cite{armani}, with the major radius $R$ and the minor radius $%
R^{\prime}$. The toroidal microcavity is then coated by a layer of polymer
material, such as SU-8, which is transparent over visible to infrared \cite%
{su8}, and has a large refraction index $n_{2}$. The coating thickness is a
nanometer scaling, denoted by $t$. Beyond the coated polymer, it is the air
with refraction index $n_{3}$. Recently, this type of polymer coated
microcavities has been utilized for enhancing the sensitivity of a
cavity-based biosensor \cite{arnold}, suppressing the thermal-optic noise in
silica resonator \cite{lina}, and observing electromagnetically induced
transparency-like effect in a single cavity \cite{xiao}. In this paper,
however, we focus on the cavity QED with coated microcavity. In the
following study, we vary the major radius $R$ and the coating thickness $t$,
but fix the size of minor radius $R^{\prime}$ for simplicity.

In order to investigate the properties of coated microtoroids for cavity
QED, we focus our attention on the dipole transition with a zero-phonon line
at $637$ \textrm{nm} of the nitrogen-vacancy (NV) defect in diamond
nanocrystal. NV centers are well suited for cavity QED and quantum
information because they have a long-lived spin triplet (over $0.35$ \textrm{%
ms}) in its electronic ground state \cite{kennedy,gaebel}\ that can be
initialized, manipulated, and read out through highly stable optical and
microwave excitations at room temperature. Experimentally, single NV centers
have been studied as a photostable single-photon source without
photobleaching for quantum communication \cite{sps1,sps2}, quantum register,
Rabi oscillations and conditional two-qubit quantum gates for quantum
information and computation \cite{nv1,nv2,nv3,nv4,nv5}. Single NV center
strongly coupling to a cavity mode has also been demonstrated \cite%
{nv6,park,benson}. In essence, the investigation of the coupling between a
NV center and a cavity mode field can be specified by three cavity-QED
parameters: the quality factors $Q$ of the cavity mode, the mode volume $V_{%
\mathrm{m}}$, and the effective dipole-cavity coupling coefficient $g(\vec{r}%
)$ at the NV center's location $\vec{r}$. In whispering gallery cavity QED
experiment, the NV centers typically interact only with the evanescent field
of the cavity mode. Thus, we assume that single NV centers are located near
the microcavity surface.

\begin{figure}[ptbh]
\centerline{\includegraphics[keepaspectratio=true,width=0.45%
\textwidth]{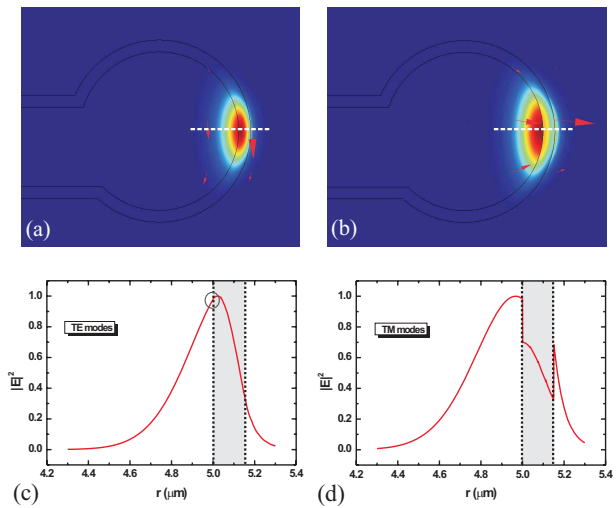}}
\caption{(Color online) (a)-(b): False-color representationes of the
normalized squared transverse electric fields for quasi-TE and TM modes,
respectively, where the arrows show the directions of the electric fields.
(c)-(d): The distributions of the normalized squared electric fields along
the radial direction (dashed lines in (a) and (b)) for quasi-TE and TM
modes, respectively. The dotted lines denote the boundary positions. Here $%
R=4$ $\mathrm{\protect\mu m}$, $R^{\prime }=1$ $\mathrm{\protect\mu m}$, and 
$t=150$ \textrm{nm}.}
\end{figure}

Different from a spherical microcavity, there is no analytic solution for a
microtoroid though the perturbative analytic theory has been proposed \cite%
{min}. For a nanolayer coated microtoroid, the perturbative analytic theory
may still work but it should be much more complicated. Therefore, to
explicitly achieve the useful metrics of the coated toroidal microcavity,
such as resonant wavelengths, quality factors and mode volumes, we resort to
numerical simulations. In Figs. 2a and 2b, by using a full-vectorial finite
element method (FEM), we show two typical WGMs in a coated silica
microtoroid in the wavelength of $637$ \textrm{nm}. Here the major and minor
radii of the silica toroid are set as $4$ and $1$ $\mathrm{\mu}$\textrm{m},
respectively, and the coating thickness is $150$ \textrm{nm}. Similar to an
uncoated microtoroid, the mode field in the coated toroidal microcavity can
be treated as quasi-TE or quasi-TM modes. For quasi-TE modes, the dominant
electric field components are in the azimuthal vertical direction, as shown
by the red arrows in Fig. 2a; while for quasi-TM modes, the dominant
electric fields are in the radial direction, as shown in Fig. 2b. Their
normalized squared transverse electric fields are also plotted in Figs. 2c
and 2d, respectively. It can be found two important indications. On one
hand, for quasi-TE modes, the electric field shows a relatively smooth
boundary change along the radial direction since it dominates in the
azimuthal vertical (transverse) direction which has continuous nature of
dielectric boundary $E_{i,\mathrm{tran}}=E_{j,\mathrm{tran}}$. Nevertheless,
there is some minor discontinuty at the boundary interface, as indicated in
the ellipse in Fig. 2c. The reason is that the toroidal geometry has been
strongly compressed with respect to a microsphere, and strictly speaking,
the quasi-TE modes are hybrid modes. As a result, the nontransverse electric
field does exist though they remain relatively small. For quasi-TM modes, as
expected, a large discontinuty of the electric fields on the boundary
occurs. This is because the dominant electric field component of the
quasi-TM mode is in the radial direction (normal to the boundary interface)
which satisfies the discontinous boundary condition $n_{i}^{2}E_{i,\mathrm{%
normal}}=n_{j}^{2}E_{j,\mathrm{normal}}$. On the other hand, in the presence
of the nanolayer coating, the maximum of the electric field of the WGMs
further move to the silica-polymer boundary interface compared to that of an
uncoated silica toroidal microcavity. For quasi-TE modes, the maximum can
even be located in the polymer layer for a smaller coating thickness. This
qualitatively demonstrates the first motivation of this paper that the
effective dipole-cavity coupling rate when the NV center is placed the outer
surface of the silica can be increased for a coated microcavity. In the
following sections, we in turn discuss the quality factors $Q$ and the mode
volume $V_{\mathrm{m}}$ of the WGMs in the coated silica microtoroid, and
the effective dipole-cavity coupling rate $g(\vec
{r}=R+R^{\prime})$. Also,
we will only consider the fundamental modes for both the
quasi-polarizations, as they possess the smallest mode volumes and thus the
largest coupling strengths.

\begin{figure}[ptbh]
\centerline{\includegraphics[keepaspectratio=true,width=0.45%
\textwidth]{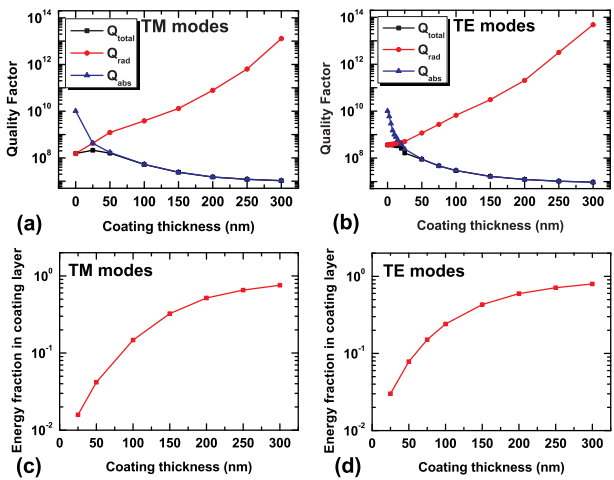}}
\caption{(Color online) (a)-(b): Quality factors for the fundamental
quasi-TM/TE WGMs in polymer coated silica microtoroids when the coating
thickness $t$ is changed. The square, circle and triangle represent the
total, radiation, and absorption related quality factors, respectively.
(c)-(d): The energy fractions in the coating polymer depending on the
coating thickness for TM and TE modes, respectively. Here $R=4$ $\mathrm{%
\protect\mu m}$ and $R^{\prime}=1$ $\mathrm{\protect\mu m}$.}
\end{figure}

\section{Quality factors of the modes}

For cavity QED experiment, the quality factor of the WGM is one of important
parameters. It is related to the several different optical loss mechanisms.
In general, the overall quality factor can be calculated by adding the
different contributions in the following express%
\begin{equation}
\frac{1}{Q_{\mathrm{total}}}=\frac{1}{Q_{\mathrm{rad}}}+\frac{1}{Q_{\mathrm{%
mat}}},  \label{eq1}
\end{equation}
and%
\begin{equation}
\frac{1}{Q_{\mathrm{mat}}}=\frac{1}{Q_{\mathrm{s.s}}}+\frac{1}{Q_{\mathrm{w}}%
}+\frac{1}{Q_{\mathrm{abs}}},  \label{eq2}
\end{equation}
where $Q_{\mathrm{rad}}$ is due to purely radiative losses for an ideal
dielectric microcavity; $Q_{\mathrm{mat}}$ results from non-ideal material
properties, including the contributions the scattering losses from residual
surface inhomogeneities ($Q_{\mathrm{s.s}}$), absorption losses due to water
on the outer surface of the coating polymer ($Q_{\mathrm{w}}$), and bulk
absorption in the silica and polymer material ($Q_{\mathrm{abs}}$).
Typically, the silica toroid has an atom-scale surface due to its
laser-reflow fabrication, and this smooth surface will transfer to the
polymer, resulting a very small scattering losses from residual surface
inhomogeneities; on the other hand, the water absorption can be suppressed
in a vacuum environment. Thus here we consider only radiation-related
quality factor $Q_{\mathrm{rad}}$, material absorption-related quality
factor $Q_{\mathrm{abs}}$, and the total quality factor $Q_{\mathrm{total}}$%
. Since in this paper we focus on the dipole transition of NV defect at
center wavelength $637$ \textrm{nm}, we set the absorption losses of silica
and SU-8 $10^{-5}$ and $0.01$ \textrm{dB/cm}, respectively.

\begin{figure}[ptbh]
\centerline{\includegraphics[keepaspectratio=true,width=0.45%
\textwidth]{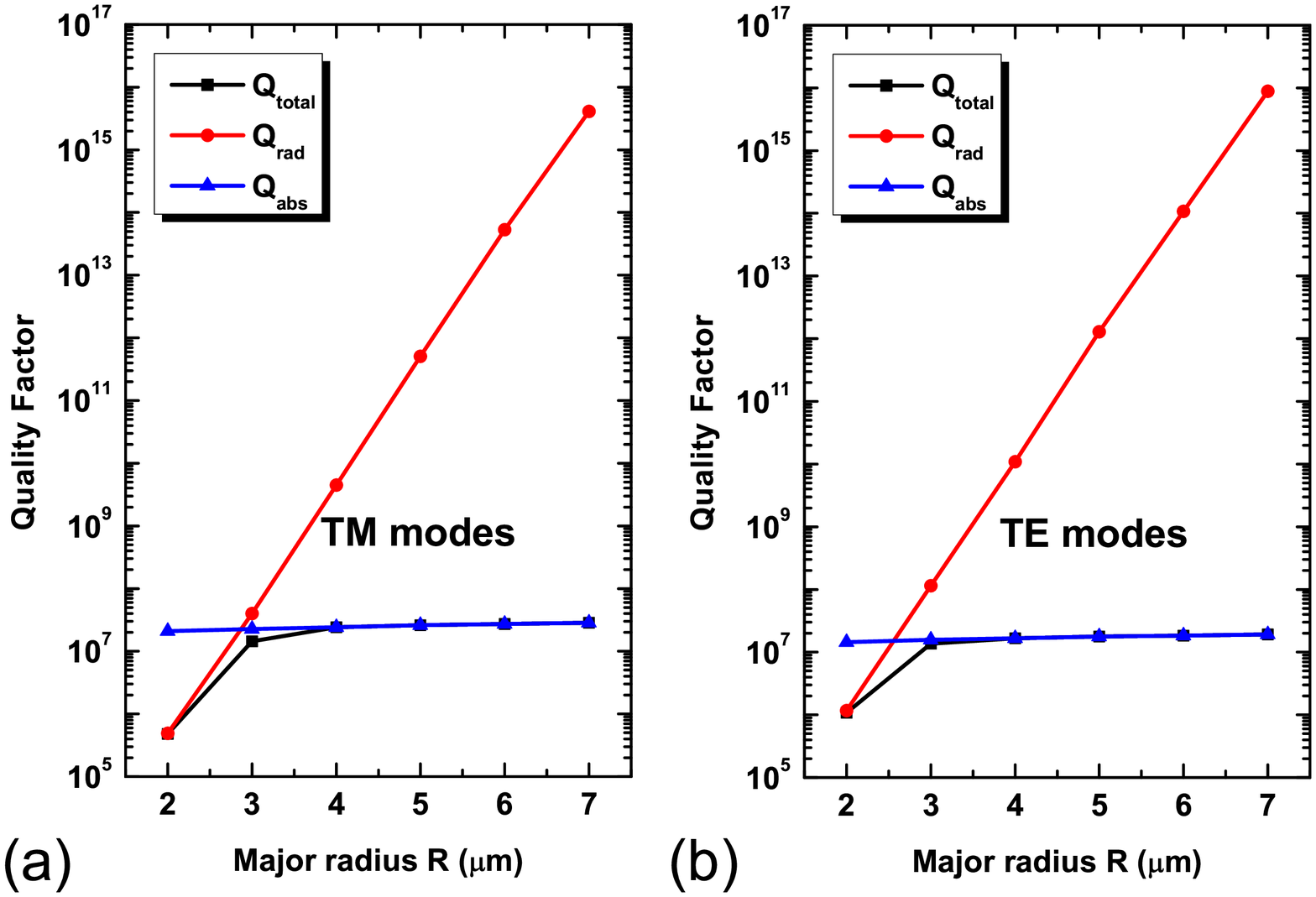}}
\caption{(Color online) (a)-(b): Quality factors for the fundamental
quasi-TM/TE WGMs in polymer coated silica microtoroids when the major radius 
$R$ is changed. The square, circle and triangle represent the total,
radiation, and absorption related quality factors, respectively. (c)-(d):
The energy fractions in the coating polymer depending on the coating
thickness for TM and TE modes, respectively. Here $t=150$ $\mathrm{nm}$ and $%
R^{\prime}=1$ $\mathrm{\protect\mu m}$.}
\end{figure}

In order to study how the quality factors depend on the coating thickness $t$%
, Figs. 3a and 3b illustrate $Q_{\mathrm{total}}$, $Q_{\mathrm{rad}}$, and $%
Q_{\mathrm{abs}}$ for quasi-TM and TE modes, respectively. For an uncoated
silica toroid ($t=0$ \textrm{nm}), the total quality factor $Q_{\mathrm{total%
}}$ behaves the radiation loss limited (i.e., $Q_{\mathrm{total}}\sim Q_{%
\mathrm{rad}}$) since the absorption of silica is much smaller (resulting $%
Q_{\mathrm{abs}}>10^{10}$) for both the TM and TE modes. This is valid if we
neglect the surface scattering loss for a small sized microsphere. For the
coated silica toroid with a large coating thickness, for example $t>50$ 
\textrm{nm}, however, $Q_{\mathrm{total}}$ is limited by the absorption loss
(thus $Q_{\mathrm{total}}\sim Q_{\mathrm{abs}}$) since in this case the
absorption of polymer plays the dominant role in the whole loss mechanisms.
This can be partly demonstrated in Figs. 3c and 3d. With increasing the
coating thickness, more and more energy distributes in the nanolayer. When
the thickness $t$ ranges from $0$ and $50$ \textrm{nm}, it is a little
complicated. There is a tradeoff between $Q_{\mathrm{rad}}$ and $Q_{\mathrm{%
abs}}$. In general, with increasing the thickness $t$, the radiation loss
rapidly decreases due to the\ protection of the high-index nanolayer
(leading to a quick increasing of $Q_{\mathrm{rad}}$), while the absorption
loss increases because more mode energy distributes in the relatively high
loss polymer layer (resulting in a decreasing of $Q_{\mathrm{abs}}$). For
both the modes, when the coating thickness $t$ increases from $0$, the total
quality factor $Q_{\mathrm{total}}$ first shows a slight increase, because
the increasing rate of the absorption loss is less than the decreasing rate
of the radiation loss. At a certain thickness $t$ ($\sim25$ and $5$ \textrm{%
nm} for quasi-TM and TE modes, respectively), $Q_{\mathrm{total}}$ reaches
the maximum when the increasing rate of the absorption loss is right
compensated by the decreasing rate of the radiation loss. Then with the
coating thickness increases further, $Q_{\mathrm{total}}$ decreases because
the increasing rate of the absorption loss exceeds the decreasing rate of
the radiation loss.

Quality factors are also dependent on the the size of toroid. Here we give
the coating thickness $t=150$ \textrm{nm} and the minor radius $R^{\prime}=1$
$\mathrm{\mu m}$, and study their dependence on the major radius $R$, as
shown in Figs. 4a and 4b, for quasi-TM and TE modes, respectively. We have
the following results. First, the absorption related quality $Q_{\mathrm{abs}%
}$ only experiences slight increase with the major radius $R$ changing.
Second, at small major radius, the radiation loss is much higher than the
absorption loss, resulting in the radiation-limited total quality factor $Q_{%
\mathrm{total}}$ ($Q_{\mathrm{total}}\sim Q_{\mathrm{rad}}$). Third, when
the major radius $R$ increases, the radiation loss decreases and becomes
comparable with the absorption loss. In this case, $Q_{\mathrm{total}}$
increases fast with increasing $R$. Finally, when $R$ is large enough, the
radiation loss becomes much lower than the absorption loss, resulting in the
absorption-limited $Q_{\mathrm{total}}$ ($Q_{\mathrm{total}}\sim Q_{\mathrm{%
abs}}$).

\begin{figure}[ptbh]
\centerline{\includegraphics[keepaspectratio=true,width=0.45%
\textwidth]{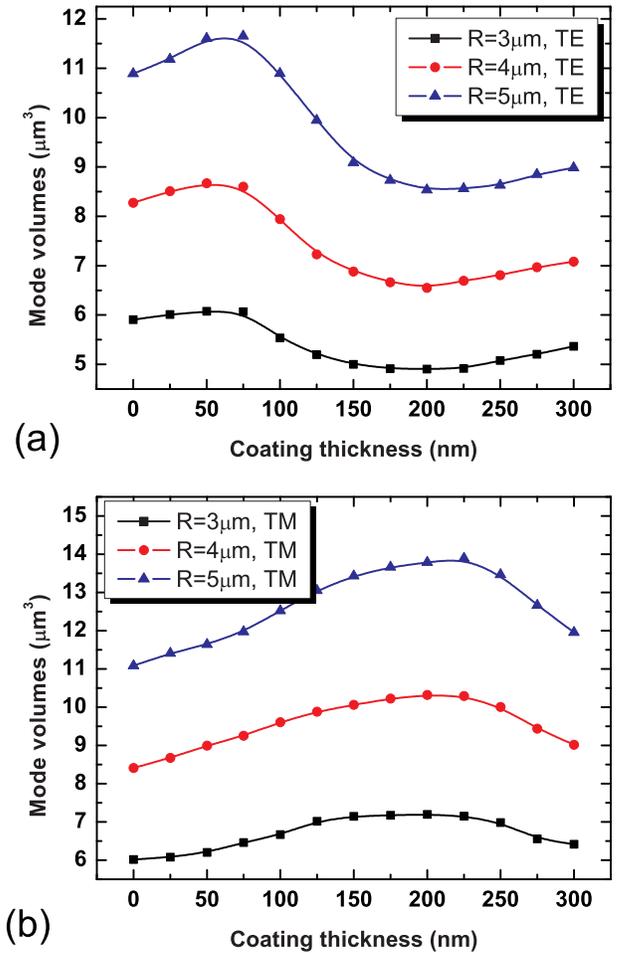}}
\caption{(Color online) (a)-(b): The mode volumes for quasi-TE and TM modes,
respectively. Here $R^{\prime}=1$ $\mathrm{\protect\mu m}$.}
\end{figure}

\section{Modal volumes of the modes}

The other important parameter for cavity QED is the modal volume $V_{\mathrm{%
m}}$ of WGM, defined as%
\begin{equation}
V_{\mathrm{m}}\equiv\frac{\int_{V}\left\vert n(\vec{r})E(\vec{r})\right\vert
^{2}dV}{\left\vert n(\vec{r})E(\vec{r})\right\vert _{\max}^{2}}.  \label{eq3}
\end{equation}
Using Eq. (\ref{eq3}), the modal volume is numerically calculated in Fig. 5
for different coating thickness and major radius. The mode volume may
increase or decrease at different coating thickness. To better explain this
phenomenon, we note that the mode volume also could be approximately
expressed as $2\pi R_{\mathrm{eff}}A_{\mathrm{eff}}$, where $R_{\mathrm{eff}%
} $ is the effective radius of the ring-type cavity, and $A_{\mathrm{eff}}$
represents the effective mode area in the cross section (see Figs. 1a and 1b
showing the cross sectional distributions for quasi-TE and TM mode fields).
Typically, there are the following three regions. (i) When the SU-8 polymer
coating thickness on the silica microtoroid is small, the mode volume first
exhibits an obvious increasing with $t$ for both the quasi-TE and TM modes.
This is due to the increase of the effective radius $R_{\mathrm{eff}}$,
approximately proportional to $R+R^{\prime}+t$, while $A_{\mathrm{eff}}$
keeps almost unchanged (minor energy in the coating nanolayer). (ii) With
the coating thickness increasing, it is interesting that $V_{\mathrm{m}}$
starts to decrease. This can be understood by considering the fact that the
refractive index $n_{2}$ of SU-8 polymer is larger than the silica. The
coating polymer film forms a kind of film waveguide, where light could be
well confined in the film by the total internal reflection at both the
polymer-silica and polymer-silica interfaces. The guiding modes in film must
satisfy the cut-off conditions. As a result, when $t$ is as large as $200$ 
\textrm{nm}, the most field energy of the mode could be well confined in the
film, resulting in the reduced mode area $A_{\mathrm{eff}}$. The decreasing
rate of $A_{\mathrm{eff}}$ exceeds the increasing rate of $R_{\mathrm{eff}}$%
, so that the mode volume decreases in a range of $t$. (iii) Finally, with $%
t $ keeping increased, the polymer-silica interface plays a minor role in
confining light, so the mode volume increases again with $t$ (The decreasing
rate of $A_{\mathrm{eff}}$ is less the increasing rate of $R_{\mathrm{eff}}$
again). The change of the mode volumes for quasi-TE and TM polarized modes
are slightly different, originating from the different cut-off condition of
guide mode for different polarization.

\begin{figure}[ptbh]
\centerline{\includegraphics[keepaspectratio=true,width=0.45%
\textwidth]{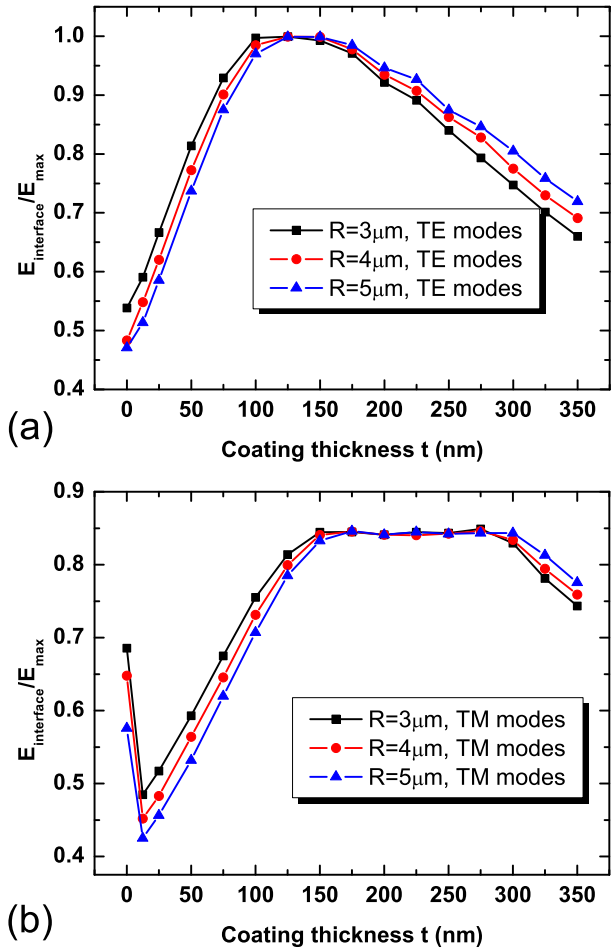}}
\caption{(Color online) (a)-(b): Electric field at silica-polymer outer
interface normalized to the maximum for quasi-TE and TM modes, respectively.
Here $R^{\prime}=1$ $\mathrm{\protect\mu m}$.}
\end{figure}

\section{Single-photon coherent coupling strength and the strong-coupling
regime}

As we mentioned above, the ultimate goal of this paper is to employ the WGMs
of coated silica microtoroids as cavity modes for achieving strong coupling
to NV centers within the setting of cavity QED. Based on the above basic
results, we can analyze the coupling coefficient $g(\vec{r})$, which is the
coupling frequency of a NV center to a particular cavity mode and
corresponds to one-half the single-photon Rabi frequency. Since we assume
that the NV center in a diamond nanocrystal is located just at the outer
surface of the silica toroid and interacts with a WGM, we can define an
effective coupling coefficient $g_{\mathrm{eff}}$, given by%
\begin{equation}
g_{\mathrm{eff}}=f\times g_{\mathrm{\max}},  \label{eq4}
\end{equation}
where the coefficient $f\equiv\left\vert E(\vec{r}=R+R^{\prime})/E_{\max
}\right\vert $ defining the electric field at the outer silica-polymer
interface normalized to the maximum, and $g_{\mathrm{\max}}\equiv\left(
\mu^{2}\omega_{c}/2\hbar\varepsilon V_{\mathrm{m}}\right) ^{1/2}$ with $%
\mu=2.74\times10^{-29}$ $\mathrm{C\cdot m}$ representing the dipole moment
of NV center transition. In this section, we analyze $f$ and $g_{\mathrm{eff}%
}$ depending on the coating thickness, and point out the potential of
enhanced interaction between the NV center and the cavity mode in coated
microcavity.

Figure 6 depicts the coefficient $f$ when the major radius $R$ is $3$, $4$,
and $5$ $\mathrm{\mu m}$, and the coating thickness $t$ ranges from $0$ to $%
350$ nm. For the quasi-TE mode, $f$ first increases and then decreases
smoothly. The underlying physics is that the maximal electric field's
position of the quasi-TE mode is drawn to the outside from inside of the
silica toroid when the coating thickness is increasing. At a critical
thickness\textrm{\ }(for instance, about $140$ nm for $R=4$ $\mathrm{\mu m;}$%
this critical thickness slightly decreases for a smaller major radius), $f$
approaches $1$. In other words, the maximal electric field of the quasi-TE
mode in this case locates at the silica-polymer interface. For the quasi-TM
mode, it is a somewhat complicated. The coefficient $f$ suddenly decreases
in the present of the nanometer-sized polymer coating layer, because of the
discontinous boundary condition for quasi-TM modes. Then $f$ gradually
increases and decreases similar to he quasi-TE mode, with increasing the
coating thickness. Unlike the quasi-TE mode case, the maximal $f$ is about $%
0.84$, smaller than both $n_{1}/n_{2}$ and $1$. The former is due to the
presence of the non-radial electric field; while the latter is because of
the discontinous condition at silica-polymer boundary for the quasi-TM mode.
Moreover, the electric field approaches the maximum within a very large
range of the thickness $t$ because the maximal electric field is exactly on
the inner surface of the silica in this range.

\begin{figure}[ptbh]
\centerline{\includegraphics[keepaspectratio=true,width=0.45%
\textwidth]{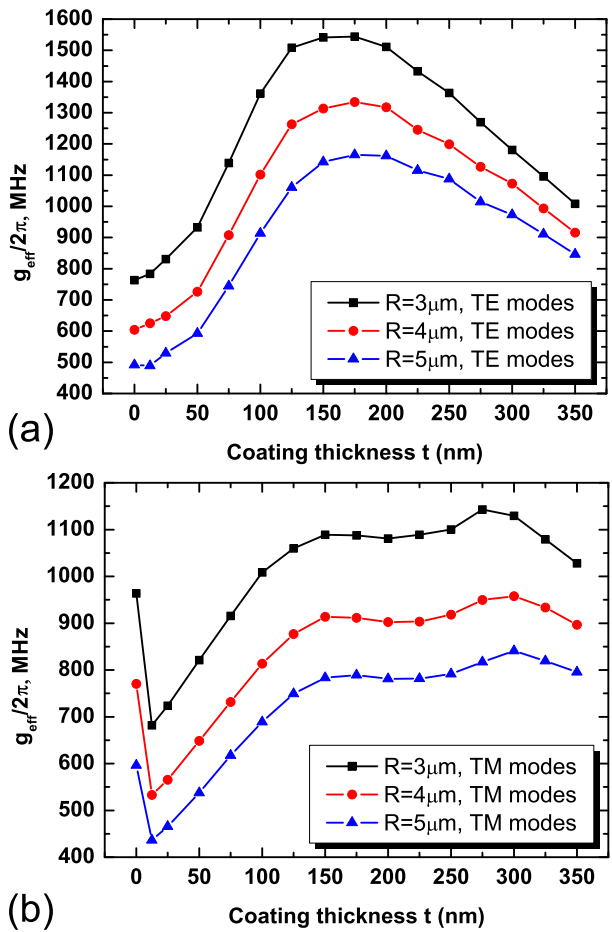}}
\caption{(Color online) (a)-(b): Effective coupling coefficient $g_{\mathrm{%
eff}}$ when a single NV center locates at silica-polymer outer interface for
quasi-TE and TM modes, respectively. Here $R^{\prime}=1$ $\mathrm{\protect%
\mu m}$.}
\end{figure}

By now, we have obtained the mode volume $V_{\mathrm{m}}$ and the
coefficient $f$. Thus, we calculate the effective coupling coefficient $g_{%
\mathrm{eff}}$ according to Eq. (\ref{eq4}), as shown in Fig. 7 for both the
quasi-TE and TM modes. Since the coating nanolayer moves the maximum of the
mode field to the silica-polymer interface, the effective coupling
coefficient $g_{\mathrm{eff}}$ is greatly enhanced. For example, $g_{\mathrm{%
eff}}/2\pi$ is of the order of \textrm{GHz}, and exceeds twice of the
uncoated case for the quasi-TE mode. Thus, an enhanced interaction can be
achieved between the WGM in the polymer-coated silica microtoroid and the
coupled NV center.

We now turn to compare the effective coupling coefficient $g_{\mathrm{eff}}$%
, the cavity mode decay rate $\kappa=\omega_{c}/Q$, and the transverse
spontaneous emission rate $\gamma/2\pi=13$ \textrm{MHz} for the dipole
transition of NV center. For an intuitive comparison, we give an example
showing quantitative values of them. For an polymer coated silica toroid
with the parameters $R=4$ $\mathrm{\mu m}$, $R^{\prime}=1$ $\mathrm{\mu m}$, 
$n_{1}=1.4564$, $n_{2}=1.59$, $n_{3}=1$, and the coating thickness $t=100$ 
\textrm{nm}, the effective coupling coefficient $g_{\mathrm{eff,TE}}/2\pi$ $%
\sim$ $1.1$ \textrm{GHz}, the cavity decay rate $\kappa_{\mathrm{TE}%
}/2\pi\sim$ $16$ \textrm{MHz} for a quasi-TE mode. Therefore, we can easily
know that the coated silica toroid provides a good platform to achieve
strong coupling regime because of $g_{\mathrm{eff}}\gg\kappa,\gamma$.

\section{Discussions and experimental feasibility}

In the section above, we have theoretically demonstrated the enhanced NV
center-WGM interaction for an coated silica microcavity. One may argue that
the coating layer may also reduce the total quality factor $Q_{\mathrm{total}%
}$\ of the cavity mode (as shown in Fig. 3), as well as an actually degraded 
$g_{\mathrm{eff}}/\kappa$ though $g_{\mathrm{eff}}$ can be improved. Thus
more discussions on the coated microcavity are necessary. First, as pointed
out in the section III, the total quality factor $Q_{\mathrm{total}}$ is
limited by the SU-8 polymer absorption loss when the coating layer exceeds a
certain thickness. As a result, $Q_{\mathrm{total}}$ has already reduced
before the coupling system reaches the maximal coherent interaction (i.e., $%
f=f_{\max}$). To exploit the enhanced-interaction advantages of the coated
microcavity, SU-8 polymer coating is particularly suitable for a smaller
silica toroid. For example, when the major radius $R=2$ $\mathrm{\mu m}$, $%
Q_{\mathrm{total,TE}}\sim10^{6}$ is still limited by the radiation losses
even in the presence of the coating layer ($t=150$ $\mathrm{nm}$). In this
case, $g_{\mathrm{eff}}$ is strongly enhanced while $\kappa$ keeps
unchanged. Second, the reduction of $\kappa$ can be relaxed by choosing a
lower loss polymer, such as PMMA, leading to an almost unaffected $Q_{%
\mathrm{total}}$ (high up to $10^{8}$) in a PMMA-coated silica microsphere 
\cite{dong}. Moreover, the refraction index of PMMA $n_{\mathrm{PMMA}}$ is
about $1.49$, which is closed to that of silica. Thus, not only the
coefficient $f_{\mathrm{TE}}$ of the quasi-TE mode is increasing, but also $%
f_{\mathrm{TE}}$ can be further improved to $n_{\mathrm{1}}/n_{\mathrm{PMMA}%
}=0.97$. Third, in an actual WGM cavity QED experiment, the deposition of
quantum dots or diamond nanocrystals on the silica surface will inevitably
degrade the total quality factor due to additional scattering losses,
particularly for the case when silica cavity is immersed in a solution and
then withdrawn \cite{park,minAPL}. The existence of the polymer coating
nanolayer, however, could recover the high total quality factor because the
polymer nanolayer has a much more smooth surface which strongly reduce the
scattering losses. Finally, compared to the conventional case that single
atoms or nanocrystals are just adsorbed on the silica surface, the polymer
nanolayer provides a protection, and single nanocrystals can stably locate
on the silica surface.

It is also noted that the spontaneous emission rate given above is the value
in the vacuum. When NV centers are placed in a dielectric, this rate will be
changed, and can be approximately calculated as $\gamma =9\varepsilon
^{5/2}/(2\varepsilon +\varepsilon _{\mathrm{NV}})^{2}\gamma _{\mathrm{vac}}$
where $\varepsilon $ and $\varepsilon _{\mathrm{NV}}$ are dielectric
constants for surrounding medium and diamond \cite{Gibbs}, respectively. In
the present paper, it is much more complex since the surrounding medium
includes both silica and SU-8 polymer. Nevertheless, the difference is
estimated about $50\%$ if we take the average value for silica and SU-8
polymer. This change may partly reduce the cavity QED parameter $g_{\mathrm{%
eff}}/\kappa \gamma $. The experimental realization of the polymer coating
on the silica surface is the essential requirement in this study.
Fortunately, polymer generally has a low surface tension enabling it to wet
many materials including silica. Polymer-coated silica microcavities have
been experimentally demonstrated recently \cite{lina,arnold}, and atomic
force microscope scans reveal that the rms roughness of the polymer surface
is comparable to that of bare silica. This makes our design has a highly
experimental feasibility.

\section{Summary}

We have theoretically studied both the quasi-TE and TM fundamental
whispering gallery modes in polymer coated silica microtoroids. Both the
modes show very high quality factors and small mode volumes. It is of
importance that the coated silica microcavity strongly enhances the coherent
interaction strength at least twice between the WGM and the coupled NV
center, compared to the uncoated microcavity, because (1) the coating layer
(though nanometer sized) can draw the maximal electric field of the mode to
the outside of the silica toroid where NV centers are located; (2) the mode
volume can be reduced due to the high refraction index polymer layer.
Furthermore, the coating nanolayer makes the cavity QED system more robust.
Thus, the polymer coated microtoroid is highly feasible in experiment and
offers a good platform to study strong-coupling cavity QED, quantum
information and quantum computation.

\begin{acknowledgments}
The authors acknowledge financial support from the National Natural Science
Foundation of China under Grant No. 10821062, the National Basic Research
Program of China under Grant Nos. 2006CB921601, 2007CB307001 and
2009CB930504. Yun-Feng Xiao was also supported by the Research Fund for the
Doctoral Program of Higher Education (No.20090001120004) and the Scientific
Research Foundation for the Returned Overseas Chinese Scholars. Chang-Ling
Zou, Chun-Hua Dong and Zheng-Fu Han were supported by the National
fundamental Research Program of China under Grant No 2006CB921900; National
Science Foundation of China under Grant No. 60537020 and 60621064. Peng Xue
was also supported by National Natural Science Foundation of China under
Grant No. 10944005.
\end{acknowledgments}

\end{document}